\documentclass[11pt]{article}
\usepackage{latexsym}
\usepackage{graphicx}
\def\ltsim{\lower3pt\hbox{$\, 
\buildrel < \over \sim \, $}}  
\def\gtsim{\lower3pt\hbox{$\, \buildrel > \over \sim \, $}}  
\makeatletter
\def\section{\@startsection {section}{1}{\z@}{-3.5ex plus -1ex minus
 -.2ex}{2.3ex plus .2ex}{\large\bf}}
\def\subsection{\@startsection{subsection}{2}{\z@}{-3.25ex plus -1ex
minus -.2ex}{1.5ex plus .2ex}{\normalsize\bf}}
\makeatother
\makeatletter
\def\theequation{\arabic{section}.\arabic{equation}}

\@addtoreset{equation}{section}
\renewcommand{\theequation}{\thesection.\arabic{equation}}
\makeatother
\newcommand{\captionfonts}{\small}
\makeatletter  % Allow the use of @ in command names
\long\def\@makecaption#1#2{%
  \vskip\abovecaptionskip
  \sbox\@tempboxa{{\captionfonts #1: #2}}%
  \ifdim \wd\@tempboxa >\hsize
    {\captionfonts #1: #2\par}
  \else
    \hbox to\hsize{\hfil\box\@tempboxa\hfil}%
  \fi
  \vskip\belowcaptionskip}
\makeatother   % Cancel the effect of \makeatletter
\catcode`@=11
\def\marginnote#1{}
\newcount\hour
\newcount\minute
\newtoks\amorpm
\hour=\time\divide\hour
by60
\minute=\time{\multiply\hour by60 \global\advance\minute
by-\hour}
\edef\standardtime{{\ifnum\hour<12 \global\amorpm={am}
\else\global\amorpm={pm}\advance\hour by-12 \fi
 \ifnum\hour=0
\hour=12 \fi
 \number\hour:\ifnum\minute<10
0\fi\number\minute\the\amorpm}}
\edef\militarytime{\number\hour:\ifnum\minute<10
0\fi\number\minute}
\def\draftlabel#1{{\@bsphack\if@filesw
{\let\thepage\relax
 \xdef\@gtempa{\write\@auxout{\string
\newlabel{#1}{{\@currentlabel}{\thepage}}}}}\@gtempa
 \if@nobreak
\ifvmode\nobreak\fi\fi\fi\@esphack}
\gdef\@eqnlabel{#1}}
\def\@eqnlabel{}
\def\@vacuum{}
\def\draftmarginnote#1{\marginpar{\raggedright\scriptsize\tt#1}}
\def\draft{\oddsidemargin
0.0truein
 \def\@oddfoot{\sl preliminary draft \hfil
\rm\thepage\hfil\sl\today\quad\militarytime}
 \let\@evenfoot\@oddfoot
\overfullrule 3pt
 \let\label=\draftlabel
\let\marginnote=\draftmarginnote
\def\@eqnnum{(\theequation)\rlap{\kern\marginparsep\tt\@eqnlabel}
\global\let\@eqnlabel\@vacuum}
}
\catcode`@=12
\def\XXint#1#2#3{{\setbox0=\hbox{$#1{#2#3}{\int}$}
     \vcenter{\hbox{$#2#3$}}\kern-.5\wd0}}

\def\bea{\begin{eqnarray}} \def\eea{\end{eqnarray}}

\def\simgt{\stackrel{>}{{}_\sim}}
\def\be{\begin{eqnarray}} \def\ee{\end{eqnarray}} 
\newcommand{\promille}{%
  \relax\ifmmode\promillezeichen
        \else\leavevmode\(\mathsurround=0pt\promillezeichen\)\fi}
\newcommand{\promillezeichen}{%
  \kern-.05em%
  \raise.5ex\hbox{\the\scriptfont0 0}%
  \kern-.15em/\kern-.15em%
  \lower.25ex\hbox{\the\scriptfont0 00}}
\begin{document}

\thispagestyle{empty}

\begin{center}
\hfill IFT-UAM/CSIC-08-23\\
\hfill CERN-PH-TH/2008-089\\
\hfill UAB-FT-644

\begin{center}

\vspace{1.7cm}

{\LARGE\bf Phantom Higgs from Unparticles}

\end{center}

\vspace{0.7cm}

{\bf A. Delgado$^{\,a}$, J. R. Espinosa$^{\,b,c}$, J. M. 
No$^{\,b}$ and M. Quir\'os$^{\,d}$}\\

\vspace{1.2cm}

${}^a\!\!$
{\em {Department of Physics, 225 Nieuwland Science Hall, U. of Notre 
Dame,\\ Notre Dame, IN 46556-5670, USA}}

${}^b\!\!$
{\em { IFT-UAM/CSIC, Fac. Ciencias UAM, 28049 Madrid, Spain}}

${}^c\!\!$
{\em { CERN, Theory Division, CH-1211, Geneva 23, Switzerland}}

${}^d\!\!$
{\em { Instituci\`o Catalana de Recerca i Estudis Avan\c{c}ats (ICREA)} 
at}

{\em {IFAE, Universitat Aut{\`o}noma de Barcelona,
08193 Bellaterra, Barcelona, Spain
}}

\end{center}

\vspace{0.8cm}

\centerline{\bf Abstract}
\vspace{2 mm}
\begin{quote}\small
A renormalizable coupling between the Higgs and a scalar unparticle
operator ${\cal O}_U$ of non-integer dimension $d_U<2$ gives rise, after
electroweak symmetry breaking, to a mass gap in the unparticle continuum
and a shift in the original Higgs mass, which can end up above or below 
the
mass gap. We show that,  besides the displaced Higgs
state, a new isolated state can generically
appear in the spectrum near or below the mass gap. 
Such state (which we call phantom Higgs) is a mixture of Higgs and
unparticles and therefore has universally reduced couplings to fermions
and gauge bosons. This phenomenon could cause the mass of the lightest 
Higgs state accessible to colliders to be much smaller than the 
mass expected from the SM Lagrangian.
\end{quote}

\vfill

\newpage
\section{Introduction}

It has been recently emphasized that the Standard Model (SM) Higgs boson
can act as a privileged portal \cite{portal} to hidden sectors beyond the
SM. For the case of hidden sectors made of unparticles \cite{Georgi} (i.e.
conformally invariant sectors) this role of the Higgs boson has been
explored in some detail in \cite{DEQ,DENQ}. More specifically one
considers ~\cite{shirman} the renormalizable coupling ${\cal O}_U |H|^2$
between a scalar operator of unparticles ${\cal O}_U$ (of scaling
dimension $d_U$, with $1<d_U<2$) and the SM Higgs field. As discussed in
\cite{DEQ} such coupling induces a tadpole for ${\cal O}_U$ after the
breaking of the electroweak symmetry (inducing also the breaking of scale
invariance in the unparticle sector \cite{shirman}) and for $d_U<2$ the
value of the vacuum expectation value $\langle {\cal O}_U\rangle$ has an
infrared (IR) divergence. This divergence can be easily cured by
considering new interactions that induce an IR cutoff that makes 
$\langle {\cal O}_U\rangle$ finite: 
a simple additional interaction between the Higgs field and the 
unparticles was discussed in
Ref.~\cite{DEQ} while a quartic
self-interaction among unparticles
was instead considered in 
Ref.~\cite{DENQ}. One of the main implications of such
mechanisms was the appearance of a mass gap, $m_g$, of electroweak size
for the unparticle sector above which the unparticle continuum
extends\footnote{The structure of an unparticle continuum above a mass gap
has been related to a particular way of breaking scale invariance in the
AdS/CFT context in \cite{AdSCFTUn}.}. One expects such mass gap as a
generic feature of any mechanism that solves the IR problem. Clearly, the 
existence of a
mass gap has dramatic implications both for phenomenology and for
constraints on the unparticle sector.

In addition, Ref.~\cite{DEQ} showed that, after electroweak symmetry
breaking (EWSB), the Higgs field mixes with the unparticle continuum above
$m_g$ in a way reminiscent of the Fano-Anderson model \cite{FA}, familiar
in solid-state and atomic physics as a description of the mixing between a
localized state and a quasi-continuum. When the Higgs mass is below $m_g$,
the Higgs survives as an isolated state but with some unparticle admixture
that  modifies its properties. On the other hand, the unparticle
continuum above $m_g$ gets a Higgs contamination which can be crucial to
make it accessible experimentally. When the Higgs mass is above $m_g$ the
Higgs state gets subsumed into the unparticle continuum and the Higgs
width gets greatly enlarged by the unparticle mixing. Such behaviour is
similar to that found when the Higgs mixes with a quasi-continuum of
graviscalars \cite{gravis}.  In both cases, with $m_h$ above or below
$m_g$, the properties of the mixed Higgs-unparticle system can be
described quite neatly through a spectral function analysis.

In the case of the IR cure discussed in \cite{DENQ} one finds also
unparticle resonances induced by the mixing with the Higgs and reminiscent
of the plasmon excitations so common in condensed matter physics. In fact,
the structure of the unparticle squared-mass matrix is similar to the
Hamiltonian that describes different collective phenomena in several
fields of physics \cite{fano}.

The purpose of this paper is to revisit the IR cure proposed in
\cite{DEQ}. We explore in more detail the available parameter space and
find an additional interesting effect that was not discussed in
\cite{DEQ}. When one starts with a Higgs interaction eigenstate well above
the mass gap, this original Higgs resonance gets shifted in mass due to
unparticle mixing and gives rise to a broad Higgs state subsumed in the
unparticle continuum and close to the original Higgs interaction
eigenstate (as it was described above). However, if the 
Higgs-unparticle interaction is strong enough, in addition to the effect 
just described,
an unexpectedly light isolated pole near or below the mass gap can appear. 
This
pole is also a mixed Higgs-unparticle state which we call ``phantom 
Higgs", 
so that the spectrum can have two ``Higgses" which are therefore 
experimentally accessible. However, their masses and widths (especially 
those of the phantom Higgs) are very different from the corresponding 
values for the SM Lagrangian.

We organize the paper as follows: in Section~2 we briefly review the
stabilization mechanism for $\langle {\cal O}_U\rangle$ presented
originally in \cite{DEQ}. In Section~3 we explore more thoroughly the rich
parameter space available showing how the new effect mentioned above takes
place. In Section~4 we perform an spectral function analysis which
clarifies the structure of the spectrum in the new regime of interest and
its phenomenological implications. We conclude in Section~5. The appendix
contains an analytical proof of the correct normalization of the spectral
function used in Section~4.

\section{A Simple Solution to the Infrared Problem}

We start with the following scalar potential
\be
V_0=m^2 |H|^2+\lambda|H|^4+\kappa_U
|H|^2\mathcal O_{U}\ ,
\label{tree}
\ee
where the first two terms are the usual SM Higgs potential and the last
term is the Higgs-unparticle coupling ($\kappa_U$ has mass
dimension $2-d_U$). As usual, the quartic coupling $\lambda$ would be
related in the SM to the Higgs mass at tree level by $m_{h0}^2=2\lambda
v^2$ (for $m^2<0$). We write the Higgs real direction as
$Re(H^0)=(h^0+v)/\sqrt{2}$, with $v =246$ GeV.

The unparticle operator $\mathcal O_U$ has dimension $d_U$, spin zero and 
its propagator is~\cite{Georgi,Cheung}
\be
P_U(p^2)=\frac{A_{d_U}}{2\sin(\pi d_U)} 
\frac{i}{(-p^2-i\epsilon)^{2-d_U}},\quad
A_{d_U}\equiv
\frac{16\pi^{5/2}}{(2\pi)^{2d_U}}\frac{\Gamma(d_U+1/2)}
{\Gamma(d_U-1)\Gamma(2d_U)}\ .
\label{prop}
\ee

When the Higgs field gets a non zero
vacuum expectation value (VEV) the scale invariance of the
unparticle sector is broken~\cite{shirman}.  From (\ref{tree})
we see that in such non-zero Higgs background the physical
Higgs field mixes with the unparticle operator ${\cal O}_U$ and
also a tadpole appears for ${\cal O}_U$ itself
which will therefore develop a non-zero VEV.

As was done in Ref.~\cite{DEQ}, it is very convenient to use a
deconstructed version of the unparticle sector, as proposed
in~\cite{deco}. One considers an infinite tower of scalars $\varphi_n$,
($n=1,...,\infty$), with squared masses $M_n^2=\Delta^2 n$. The mass
parameter $\Delta$ is small and eventually taken to zero, limit in which
one recovers a (scale invariant) continuous mass spectrum. As explained
in~\cite{deco}, the deconstructed form of the operator ${\cal O}_U$ is
\be
\label{OUdec}
{\cal O}\equiv \sum_n F_n \varphi_n \ ,
\ee
where $F_n$ is chosen as
\be
\label{Fdec}
F_n^2 = \frac{A_{d_U}}{2\pi}\Delta^2 (M_n^2)^{d_U-2}\ , 
\ee 
so that the two-point correlator of ${\cal O}$ matches that of ${\cal
O}_U$ in the $\Delta\to 0$ limit. In the deconstructed theory then, the
unparticle scalar potential, including the coupling (\ref{tree}) to the
Higgs field, reads
\be
\label{potdec}
\delta{V} = \frac{1}{2}\sum_n M_n^2\varphi_n^2+\kappa_U |H|^2\sum_n F_n 
\varphi_n\ .
\ee
A non-zero VEV, $\langle |H|^2\rangle=v^2/2$, triggers a VEV
for the fields $\varphi_n$:
\be
\label{vn}
v_n\equiv\langle\varphi_n\rangle=-\frac{\kappa_U v^2}{2M_n^2}F_n\ ,
\ee
thus implying, in the continuum limit, 
\be
\label{vevOU}
\langle {\cal O}_U \rangle = -\frac{\kappa_U
v^2}{2} \int_0^\infty \frac{F^2(M^2)}{M^2}dM^2\ ,
\ee
where 
\be
\label{F}
F^2(M^2)=\frac{A_{d_U}}{2\pi}(M^2)^{d_U-2}\ ,
\ee
is the continuum version of (\ref{Fdec}). We see that
$\langle {\cal O}_U \rangle$ has an IR divergence for $d_U<2$, due to the
fact that for $M\to 0$ the tadpole diverges while the mass itself,
that should stabilize the unparticle VEV, goes to zero.

In Ref.~\cite{DEQ} it was shown how one can easily get an IR regulator in 
(\ref{F}) by including a coupling
\be
\label{HHUU}
\delta V = \zeta |H|^2 \sum_n \varphi_n^2\ ,
\ee
in the deconstructed theory. This coupling respects the conformal
symmetry but will break it when $H$ gets a VEV.

One can easily understand why (\ref{HHUU}) solves the IR problem in the 
continuum limit by defining the (dimensionless) field $u(x,M^2)$ by means 
of the redefinition $\varphi_n(x)=\Delta u_n(x)$ followed by 
$u(x,M^2)=\lim_{\Delta\to 0}u_n(x)$. In this way Eqs.~(\ref{potdec}) and 
(\ref{HHUU}) read as\footnote{Concerning possible problems with locality, 
note that this term satisfies the cluster decomposition principle. In the 
continuum limit this can be shown after identifying the creation operator 
for unparticles with the appropriate integral in $M^2$. In the 
deconstructed case, with a small but finite mass splitting, this principle 
is trivially satisfied.}
\begin{equation}
\delta V=\int_0^\infty dM^2\left\{\frac{1}{2} \left[M^2+2\zeta
|H|^2\right] u^2(x,M^2)+ \kappa_U|H|^2F(M^2)u(x,M^2)\right\}
\label{nueva}
\end{equation}
In the absence of the term (\ref{HHUU}) the IR problem comes from the
fact that the zero mode $u(x,0)$ is massless. However in the presence
of (\ref{HHUU}) the zero mode acquires a mass squared given by $2\zeta
|H|^2$, which in the electroweak vacuum, where conformal invariance is
broken, is given by $\zeta v^2$. In this way the term (\ref{HHUU})
introduces an IR cutoff in the theory.

Now the vacuum expectation value $\langle\mathcal O_U \rangle$ becomes
\be
\langle {\cal O}_U \rangle = -\frac{\kappa_U v^2}{2} \int_0^\infty
\frac{F^2(M^2)}{M^2+\zeta v^2}\ dM^2\ ,
\ee
where we explicitly see the presence of a mass gap at
\be
m_g^2=\zeta v^2\ ,
\ee
acting as an IR cutoff. The integral is now obviously finite for $1<d_U<2$ 
and reads explicitly
\be
\langle {\cal O}_U \rangle
=-\frac{1}{2}\kappa_U\frac{A_{d_U}}{2\pi}\zeta^{d_U-2}
v^{2d_U-2}\Gamma(d_U-1)\Gamma(2-d_U)\ .
\label{vevOUIR}
\ee
Implications for EWSB of the coupling (\ref{HHUU}) were 
studied in Ref.~\cite{DEQ}. 

\section{Exploring the Parameter Space}

In order to study the interplay between Higgs and unparticles 
we write down explicitly the infinite squared mass matrix that mixes 
the (real) neutral component $h^0$ of the Higgs with the deconstructed 
tower of unparticle scalars, $\varphi_n$.
The different matrix elements are:
\bea
\label{massmatrixhh}
M_{hh}^2 & = & 2\lambda v^2 \equiv m_{h0}^2\ ,\\
\label{massmatrixhn}
M_{h n}^2 & = & \kappa_U v F_n\frac{M_n^2}{M_n^2+m_g^2}\equiv A_n\ ,\\
M_{nm}^2 & = &  (M_n^2+m_g^2)\ \delta_{nm}\ .
\label{massmatrixnm}
\eea

It is a simple matter to obtain the $hh$-entry of the inverse (infinite 
matrix) propagator associated to this infinite mass matrix. In the 
continuum limit we obtain:
\be
\label{invprop}
iP_{hh}(p^2)^{-1}= p^2 - m_{h0}^2 + J_2(p^2)\ ,
\ee
where \cite{DEQ}
\bea
J_2(p^2)&\equiv&\int_0^\infty G_U(M^2,p^2) M^4 
dM^2=\frac{v^2}{p^4}(\mu_U^2)^{2-d_U}\Gamma(d_U-1)\Gamma(2-d_U)
\nonumber\\
&&
\times\left[\left(m_g^2-p^2\right)^{d_U}+d_U p^2 
(m_g^2)^{d_U-1}-(m_g^2)^{d_U}\right]
\label{Jk}\ ,
\eea
with
\be
G_U(M^2,p^2)\equiv\frac{v^2(\mu_U^2/M^2)^{2-d_U}}{(M^2+m_g^2-p^2)(M^2+m_g^2)^2}
\ ,
\ee
and
\be
(\mu_U^2)^{2-d_U}\equiv \kappa_U^2\frac{A_{d_U}}{2\pi}\ .
\ee

Due to the extra unparticle term in this propagator the Higgs pole will no
longer be at its SM value $m_{h^0}^2$ but displaced from it. Whether this
displacement is positive (towards higher masses) or negative will depend
on the balance between two competing eigenvalue repulsion effects: the
unparticle continuum above $m_{h^0}^2$ will tend to lower the Higgs mass
while the continuum below will tend to increase it. Of course, when
$m_{h^0}^2$ is below $m_g^2$ the shift is necessarily negative \cite{DEQ}.  
When the Higgs width is small so that one can neglect the imaginary part
of the pole (complex in general) the final outcome for the Higgs pole at
$m_h^2$ is well approximated by the solution to the pole equation
\be
{\mathcal Re}\left[iP_{hh}(m_{hR}^2)^{-1}\right]=0\ ,
\ee
where the subscript $R$ indicates that $m_{hR}^2$ is the (real) pole of 
the real part of the propagator. As discussed in \cite{DEQ}, the Higgs 
width can be greatly enlarged by unparticle mixing so that it is more 
appropriate to find the complex poles of the propagator:
\be
P_{hh}(\tilde{m}_h^2)^{-1}=0\ ,
\label{cpole}
\ee
with $\tilde{m}_h^2\equiv m_h^2-i m_h \Gamma_h$, where $m_h$ is the Higgs 
mass 
and $\Gamma_h$ the (tree-level) Higgs width.

\begin{figure}[t]
\includegraphics[width=15.cm,height=10.cm]{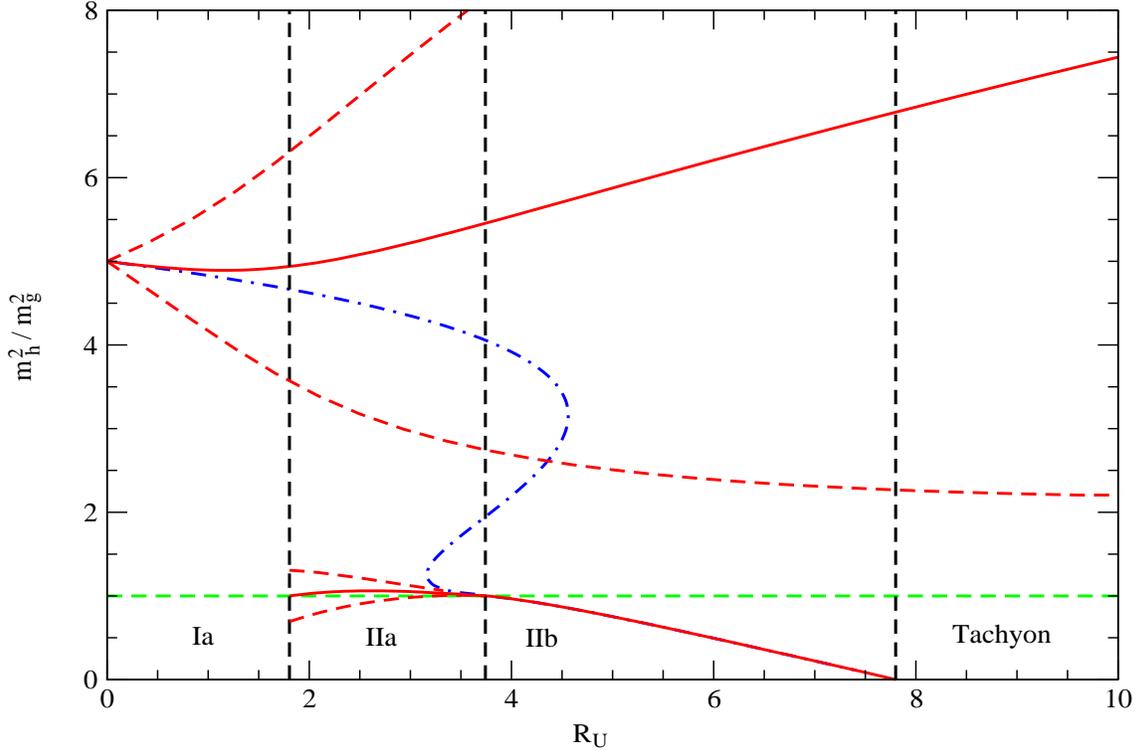}
\caption{\label{fig:mh12} The solid red curves give the Higgs pole 
masses $m_h^2$ as a function of $R_U$ for $m_{h0}^2=5m_g^2$ and 
$d_U=1.2$ while the red-dashed curves give $m_h^2\pm m_h\Gamma_h$. The 
dot-dashed blue line gives $m_{hR}$, the pole of the real part 
of the propagator. The horizontal dashed line 
gives $m_g$ and the vertical 
dashed lines delimit the different zones as indicated by the labels. }
\end{figure}

\begin{figure}[t]
\includegraphics[width=15.cm,height=10.cm]{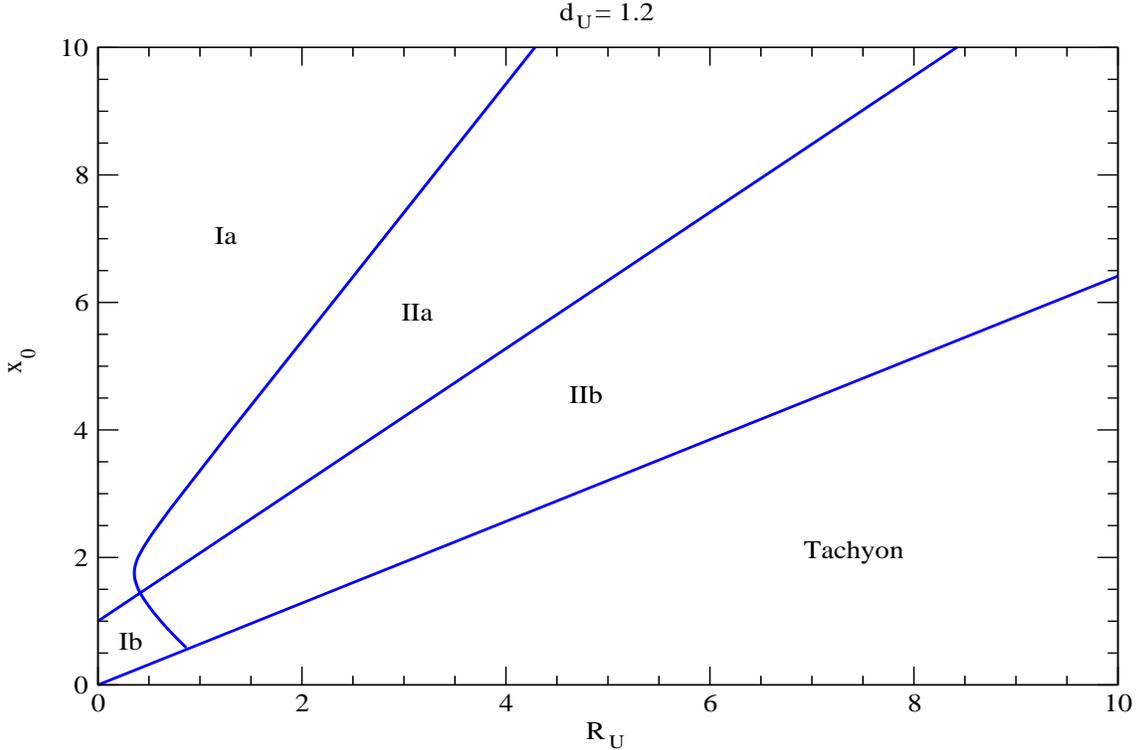}
\caption{\label{fig:zones} Different zones in the plane 
$(R_U,x_0=m_{h0}^2/m_g^2)$ with different number of 
Higgs poles: one in zone I (above $m_g$ in zone Ia, below in zone Ib) and  
two in zone II (both above $m_g$ in IIa, 
one above and one below in IIb). In the zone labeled ``Tachyon'' the 
lightest pole becomes tachyonic.} \end{figure}

In order to explore the possible qualitative behaviours of the solutions
to equation (\ref{cpole}) it is convenient to express all squared masses 
in terms of the mass gap $m_g^2$ and of the dimensionless combination
\be
R_U\equiv 
\frac{v^2}{m_g^2}
\left(\frac{\mu_U^2}{m_g^2}\right)^{2-d_U}\ ,
\ee
which measures the strength of the Higgs-unparticle interaction.
The pole equation takes then the simple form
\be
\label{pole}
\tilde{x}=x_0 -\frac{R_U}{\tilde{x}^2}f_U(\tilde{x})\ ,
\ee
where
\be
\tilde{x}\equiv \frac{\tilde{m}_h^2}{m_g^2}\ ,
\;\;\;\;\;\;\;
x_0\equiv \frac{m_{h0}^2}{m_g^2}\ ,
\ee
and 
\be
\label{fu}
f_U(\tilde{x})=\Gamma(d_U-1)\Gamma(2-d_U)
\left[(1-\tilde{x})^{d_U} + d_U \tilde{x} -1\right]\ .
\ee
In order to solve the pole equation (\ref{pole}) one should specify in
what
Riemann
sheet $z^{d_U}$ is taken in (\ref{fu}). If one sticks to the principal
sheet, with angles defined from $-\pi$ to $\pi$, the only possible poles
appear in the real axis and below the mass gap. If one goes to the second
Riemann sheet (with angles between $-3\pi$ and $-\pi$) one finds also
complex poles. We refer to these poles in the rest of the paper. The
absence of complex poles in the principal sheet will be used with
advantage in the appendix.

For small values of the unparticle effect, as measured by the parameter
$R_U$ ({\it i.e.} for $R_U\ll 1$), a perturbative solution gives
\be
m_h^2\simeq m_{h0}^2 - 
m_g^6\frac{R_U}{m_{h0}^4}{\mathcal Re}\left[f_U\left(x_0\right)\right] \ ,
\ee
with the sign of the shift determined by the sign of the function $f_U$
\cite{DEQ} and
\be
\Gamma_h \simeq m_g^6\frac{R_U}{m_{h0}^5}{\mathcal 
Im}\left[f_U\left(x_0\right)\right]\theta(x_0-1)\ .
\ee
Although the analysis of \cite{DEQ} was not restricted to very
small values of $R_U$, the behaviour of $m_h^2$ discussed there was
qualitatively similar to the one just described.

\begin{figure}[t]
\includegraphics[width=15.cm,height=10.cm]{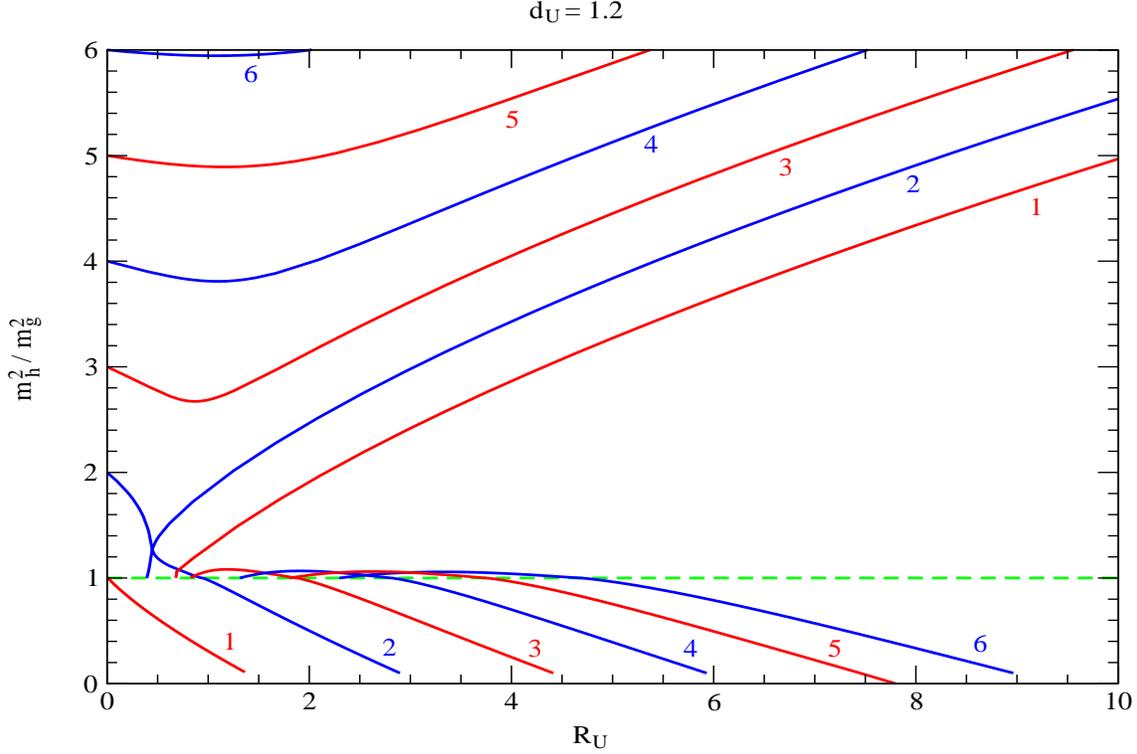}
\caption{\label{fig:MMHH} Same as Fig.~\ref{fig:mh12} for different 
values of $x_0=m_{h0}^2/m_g^2$ as indicated by the labels.} 
\end{figure}

New interesting effects occur when larger values of $R_U$ are probed.
Fig.~\ref{fig:mh12} illustrates this for the particular case $d_U=1.2$
and $m_{h0}^2/m_g^2=5$ by showing $m_h^2$ (solid lines) as a function of
$R_U$. For small $R_U$ one simply gets a negative shift for $m_h$ (zone
marked as Ia). However, for larger values of $R_U$ ($R_U\simgt 1.8$)  
things get much more interesting. In zone IIa one finds two Higgs poles
above $m_g$, one of them very close to the mass gap and the other closer
to the initial value $m_{h0}$. In zone IIb the lighter of these poles, the
phantom Higgs, goes below the mass gap while the other gets heavier.
Eventually, for sufficiently large $R_U$, the squared mass of the phantom
pole gets negative and the state becomes tachyonic. We also show the width
of these poles by giving (dashed lines) the curves for $m_h^2\pm
m_h\Gamma_h$ (we come back to the discussion of this width in section~4,
using the spectral function technique). We see that the heavy pole gets
wider and wider with increasing $R_U$ while the lighter has always a small
width. When the light Higgs gets below the mass gap its width (at
tree-level) is zero. For comparison, we also show in this figure the value
of $m_{hR}$ (dot-dashed line). We see that it approximates well $m_h$ when
the Higgs width is small but can be very different from it when the width
gets larger.

Zone II is particularly striking: the initial SM Higgs pole, which was
well above the mass gap into the unparticle continuum, gets swallowed up
by this continuum which spits out a much lighter pole near (IIa) or below
(IIb) the mass gap. A similar phenomenon has been described in other
fields of physics, see {\it e.g.} \cite{limited}. This behaviour is
generic and persists for other values of $x_0=m_{h0}^2/m_g^2$ and/or
$d_U$.  Fig.~\ref{fig:zones} shows the different zones, with the same
coding as explained above, in the plane $(x_0,R_U)$ for $d_U=1.2$. In
addition to the zones discussed above, there is also the possibility of
having a single pole below the mass gap, corresponding to zone Ib in this
plot. We do not give contour lines of $x=m_h^2/m_g^2$ as they would
overlap in regions with two poles, making the figure clumsy. Between the
lines delimiting zone Ib+IIb the mass of the pole below $m_g$ tends to 
zero
at the lower boundary (the border with the tachyonic zone) and to $m_g$ in
the upper boundary. In the boundary between zones Ia and IIa the mass of
the light Higgs is also $m_g$.

Fig.~\ref{fig:MMHH} shows $m_h^2$ vs. $R_U$ for different values of the 
initial $x_0$. The case corresponding to $x_0=1$ displays, for small 
$R_U$, 
the behaviour associated to zone Ib, with a single pole below the mass 
gap. For larger $R_U$, however, we see that an additional pole appears 
above the mass gap. Notice that (in all cases) once the lighter 
phantom Higgs 
becomes tachyonic the parameter choice is not acceptable. To answer the 
question of which pole carries a higher Higgs composition one can use a 
spectral function analysis as discussed in the next section.

\section{Spectral Function Analysis}

To clarify the pole structure of the mixed Higgs-unparticle propagator
we now turn to the study of its spectral function, given by
\be
\label{rhodef}
\rho_{hh}(s) =-\frac{1}{\pi} Im[-i P_{hh}(s+i\epsilon)]\ ,
\ee
where the limit $\epsilon\rightarrow 0$ is understood. We can calculate 
easily this spectral function by using $1/(x+i\epsilon)\rightarrow 
{\mathrm P.V.}[1/x] -i\pi\delta(x)$ directly in the integral 
$J_2$ of (\ref{Jk}) to obtain, for $s>m_g^2$,
\be
J_2(s+i\epsilon)=R_2(s)+i I_2(s)\ ,
\ee
with
\bea
R_2(s)&=&{\mathrm P.V.}[J_2(s)]\ ,\nonumber\\
I_2(s)&=&\pi 
\frac{v^2}{s^2}(\mu_U^2)^{2-d_U}(s-m_g^2)^{d_U}\ .
\eea

\begin{figure}[t]
\includegraphics[width=15.cm,height=10.cm]{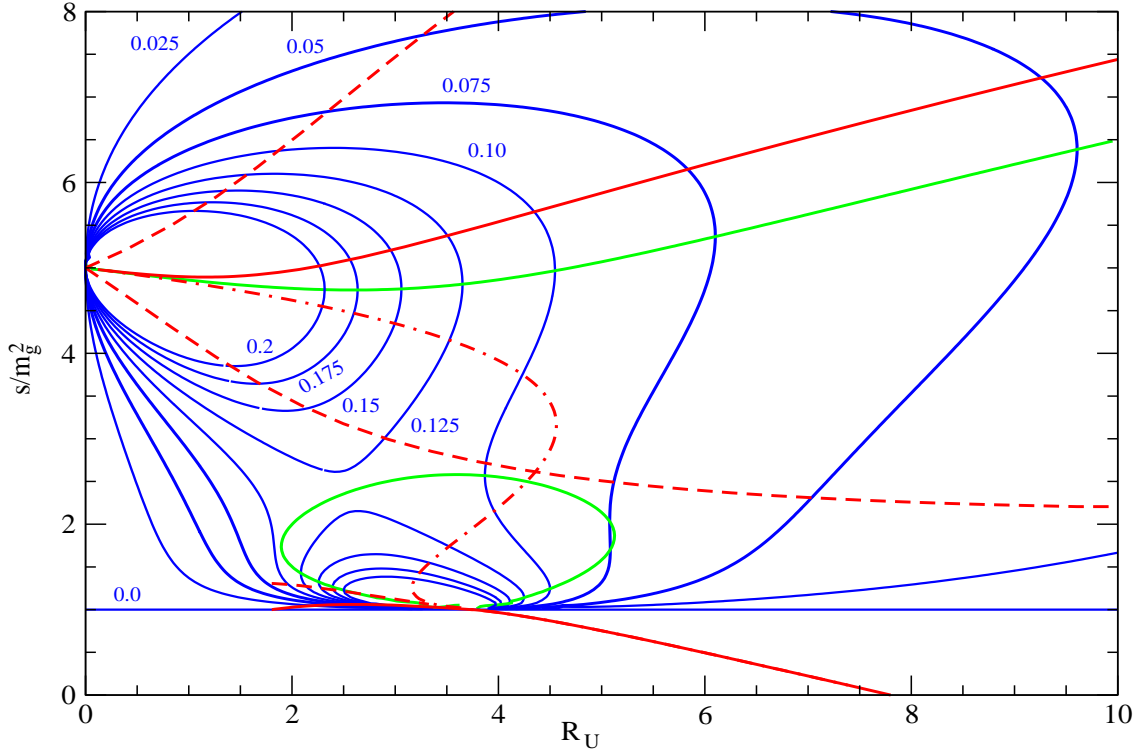}
\caption{\label{fig:contourrho} Contour lines of $\rho_{hh}(s)$ (we stop
at 0.2) in the plane $(R_U,s/m_g^2)$ for  $d_U=1.2$, $x_0=5$ (blue 
lines). Information on the Higgs poles is given by the same curves as
in Fig.~\ref{fig:mh12}. The green lines give the 
extrema of the spectral function at fixed $R_U$.} \end{figure}

\begin{figure}[t]
\includegraphics[width=15.cm,height=10.cm]{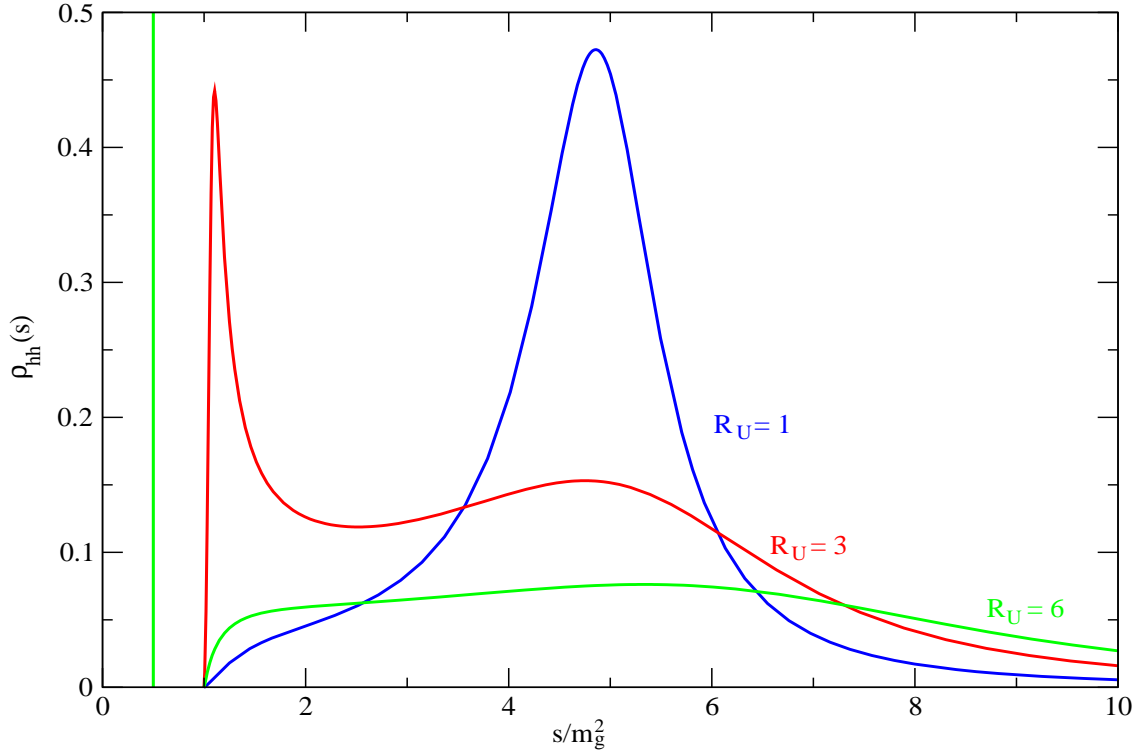}
\caption{\label{fig:rhocuts} Cuts of Fig.~\ref{fig:contourrho} along 
fixed $R_U$ values, as indicated.}
\end{figure}

When there is one pole below the mass gap, and irrespective of whether 
there is another pole above it or not, the spectral function takes the 
form \cite{DEQ}
\be
\label{rho}
\rho_{hh}(s)= \frac{1}{K^2(m_h^2)} \delta(s-m_h^2)+\theta(s-m_g^2) 
\frac{T_U(s)}{\mathcal{D}^2(s)+\pi^2T_U^2(s)}
\ ,
\ee
where $\mathcal{D}(s)$ and $\pi T_U(s)$ are the real and imaginary parts 
of 
$iP_{hh}(s+i \epsilon)^{-1}$ when $s>m_g^2$:
\be
iP_{hh}(s+i \epsilon)^{-1}=
\mathcal{D}(s)+i\pi\ T_U(s)=\left[s-m_{h0}^2 + R_2(s)\right]+ i I_2(s)\ 
.
\ee
Finally,
\be
\label{Ks0}
K^2(s_0)\equiv 
\left.\frac{d}{ds}\mathcal{D}(s)\right|_{s=s_0}\ .
\ee
An explicit expression for $K^2(m_h^2)$ can be obtained directly from 
$\mathcal{D}(s)$ above. When all the poles are above $m_g$ the spectral 
function is given by the same continuum function as in (\ref{rho}) 
without the Dirac-delta term.

One can check (see appendix for an analytical proof) that the spectral 
function (\ref{rho}) is properly
normalized:
\be
\int_0^\infty \rho_{hh}(s) ds =1\ .
\label{norm}
\ee
The physical interpretation of this spectral function was discussed in 
\cite{DENQ}: Calling $|h\rangle$ the Higgs interaction eigenstate and 
$|u,M\rangle$ the unparticle interaction eigenstates (a continuous 
function of $M$) and $|H\rangle$, $|U,M\rangle$ the respective 
mass eigenstates after EWSB, with $|H\rangle$ being the isolated state 
below the mass gap (we consider this particular case to illustrate the 
interpretation), one has
\be
\rho_{hh}(s)\equiv \langle h |s\rangle\langle s |h\rangle= |\langle H|h 
\rangle|^2 
\delta(s-m_h^2)+\theta(s-m_g^2)
|\langle U,M |h \rangle|^2
\ ,
\ee
so that one can read-off the Higgs composition of the isolated pole and
the unparticle continuum directly from (\ref{rho}). The proper
normalization (\ref{norm}) is simply a consequence of the proper
normalization of $|h \rangle$, {\em i.e.} $|\langle h|h \rangle|^2=1$.  
The amount of $|h\rangle$ admixture in any state is an important quantity
because it determines key properties of that state, like its coupling to
gauge bosons, that are crucial for its production and decay.

In Figs.~\ref{fig:contourrho} and \ref{fig:rhocuts} we show the spectral 
function for the case $d_U=1.2$, $x_0=5$ and varying $R_U$. In 
Fig.~\ref{fig:contourrho} we give contour lines of $\rho_{hh}(s)$ (we stop 
them 
at 0.2) in the plane $(R_U,s/m_g^2)$. We see two global peaks above the 
mass gap, one is at $(R_U=0,s/m_g^2=x_0)$, corresponding to the SM Higgs 
resonance, and the other at $(R_U\simeq 3.5,s/m_g^2=1)$ corresponding to 
the phantom Higgs. For $R_U\simgt 3.5$ this phantom Higgs drops below the 
mass gap giving rise to a delta pole in the spectral function. We 
show by the solid red lines the Higgs poles in this particular case 
(corresponding to Fig.~\ref{fig:mh12}). The green solid lines give the 
extrema of the spectral function for fixed $R_U$. We see that the pole 
lines offer reliable information about the location of the maxima of the 
spectral function (we should not expect perfect correspondence, see {\it 
e.g.} \cite{nucl}) and their widths
while the dashed curve corresponding to $m_{hR}$ is only a good 
approximation near the global peaks and along the isolated pole (where the 
tree-level Higgs width is small or zero). In any case, it is clear that 
the spectral function 
carries more information concerning the structure of the Higgs propagator 
than simply giving the location and width of its poles and it is therefore 
much more useful to deal directly with it. To clarify even further the 
structure of the spectral function, Fig.~\ref{fig:rhocuts} gives 
$\rho_{hh}(s)$ at various fixed values of $R_U$ for the same parameters as 
before, $d_U=1.2$ and $x_0=5$. For $R_U=1$ there is only one pole, it is 
above $m_g$ and corresponds to the somewhat wide resonance of the spectral 
function (zone Ia). One can
directly relate the width of this resonance (as measured by the width
across it at half the peak maximum) with the width as given by the dashed
lines in Fig.~\ref{fig:mh12}.
For $R_U=3$, the pole above $m_g$ has become wider and less 
pronounced while a sharper resonance has appeared right above the mass 
gap (zone IIa). Notice that the continuum part of the spectral function 
does not extend below the gap. This is in contrast with the behaviour of 
the complex pole near $m_g$ shown in Fig.~\ref{fig:mh12}: from there, 
after taking into account the width, one would conclude that the light 
resonance extends below $m_g$. For $R_U=6$ this resonance has 
detached from the continuum giving a delta function below $m_g$. The 
pole above $m_g$ is very broad 
and shallow (zone 
IIb) and could hardly be called a resonance.

\begin{figure}[t]
\includegraphics[width=15.cm,height=10.cm]{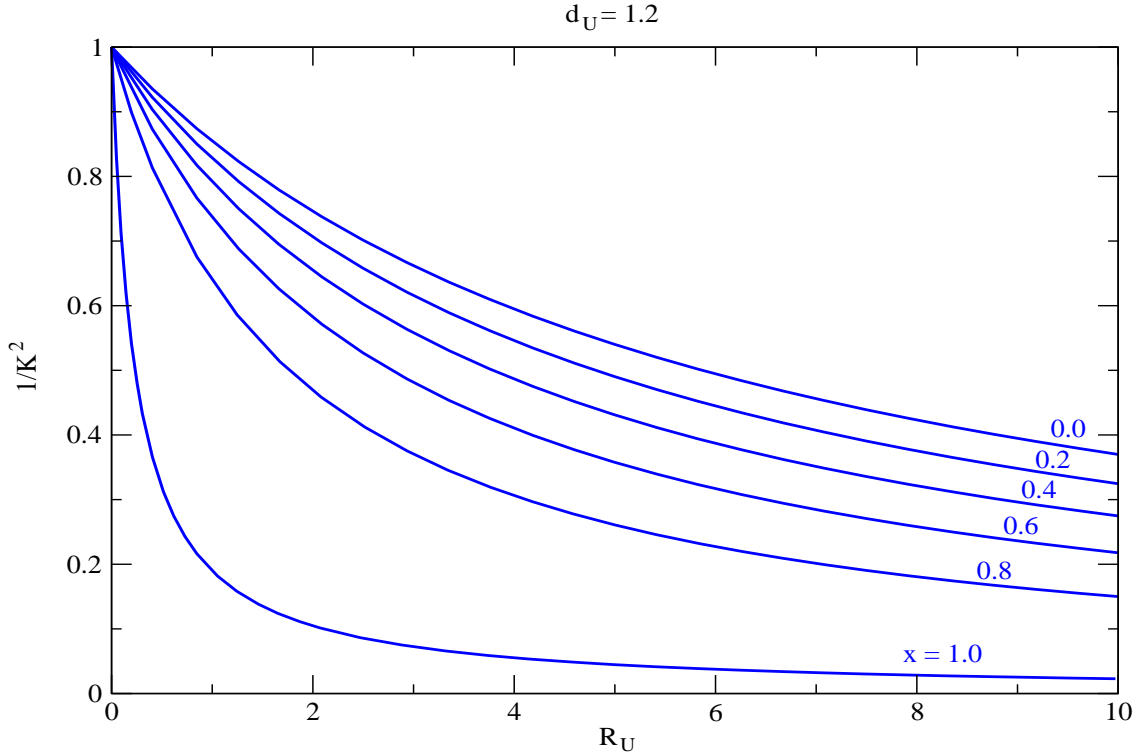}
\caption{\label{fig:K} Pure Higgs composition of the isolated pole below 
$m_g$ as a function of $R_U$ for different values of $x=m_h^2/m_g^2$ and 
for $d_U=1.2$.} 
\end{figure}

From the previous figures one cannot obtain information on the prefactor
$1/K^2$ which weights the Dirac delta contribution to  $\rho_{hh}(s)$ when 
there is a pole below $m_g$ and gives information of the pure Higgs 
composition of that pole, as explained above. This information is given by 
Fig.~\ref{fig:K} (valid for $d_U=1.2$), where the different lines 
correspond to different 
values of $x=m_h^2/m_g^2$ from $x=0$ to $x\rightarrow 1^-$. When the 
influence of 
unparticles is small 
(small $R_U$) $1/K^2\rightarrow 1$ as it should be for a Higgs with SM 
properties. The departure of $1/K^2$ from 1 is larger for larger $R_U$ 
(larger unparticle mixing) or when $m_h$ gets closer to $m_g$ (smaller 
mass difference between the states that mix).

\section{Conclusions}

The Standard Model Higgs boson offers a unique opportunity to probe the 
scalar part of an unparticle sector through a direct renormalizable 
coupling of the form $|H|^2{\cal O}_U$, where ${\cal O}_U$ is an 
unparticle scalar operator of non-integer dimension $d_U$, with $1<d_U<2$.
Several interesting effects follow from such a coupling after electroweak 
symmetry breaking, both for 
unparticle phenomenology and for Higgs boson properties, as has been 
discussed recently in \cite{shirman,DEQ,DENQ}. Among these effects we 
have: a mass gap $m_g$ of electroweak size is generated above which lies 
the unparticle continuum (which therefore does not extend all the way to 
zero mass). This unparticle continuum mixes with the Higgs so that the 
Higgs resonance gets some unparticle admixture that changes the Higgs 
couplings from its SM values while the Higgs admixture of the unparticle 
continuum helps in making it accessible experimentally. The Higgs mass is 
also affected by the unparticle mixing getting shifted from its SM value.
If it ends above the mass gap it gets subsumed in the unparticle continuum 
and becomes very wide at tree-level due to such mixing.

In this paper we have found yet another remarkable effect: starting 
with a SM Higgs mass well above the unparticle mass gap, into the 
continuum, if the Higgs-unparticle interaction is large enough, a 
``phantom'' Higgs besides the original one will appear near or below the 
mass gap. It will have some unparticle admixture and some Higgs composition 
that makes it, in principle, accessible experimentally. Therefore the 
spectrum will contain two Higgs resonances: one heavy and wide, clearly 
related to the original SM Higgs state and another thin and much lighter 
than one would naively expect from the parameters of the SM part of the 
potential.

\subsection*{Acknowledgments}

\noindent 
J.R.E. thanks CERN for hospitality and partial financial support. J.M.N. 
thanks IFAE, Barcelona, for hospitality.
Work supported in part by the European Commission under the European
Union through the Marie Curie Research and Training Networks ``Quest
for Unification" (MRTN-CT-2004-503369) and ``UniverseNet"
(MRTN-CT-2006-035863); 
by the Spanish Consolider-Ingenio 2010 Programme CPAN (CSD2007-00042);
by a Comunidad de Madrid project (P-ESP-00346)
and by CICYT, Spain, under contracts FPA 2007-60252
and FPA 2005-02211.

\section*{Appendix A}
\setcounter{equation}{0}
\renewcommand{\theequation}{A.\arabic{equation}}

\begin{figure}[t]
\includegraphics[width=10.cm,height=10.cm]{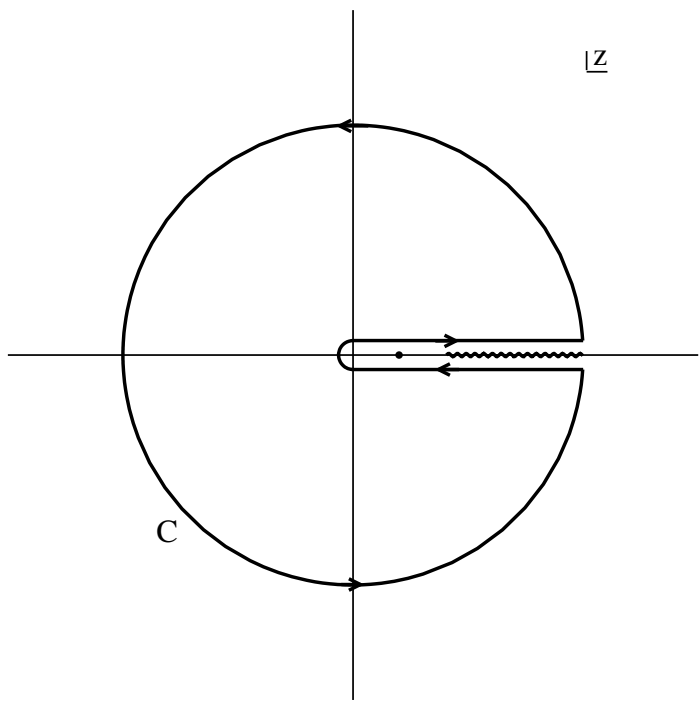}
\caption{\label{fig:contour} Integration contour for $P_{hh}(z)$ in the
complex $z$-plane.}
\end{figure}

In this appendix we give an analytical proof of the normalization
condition (\ref{norm}) for the spectral function used in section~4. The
proof uses complex integration methods very common in the
literature of dispersion techniques. Take
the $hh-$propagator of Eqs.~(\ref{invprop}-\ref{Jk}) to be defined in
the complex plane, $P_{hh}(z)$, and integrate it along the contour of
Fig.~\ref{fig:contour}, which shows the general case with a real pole
below the mass gap and a branch cut from that mass gap to infinity. The
absence
of complex poles of $P_{hh}(z)$ in the principal branch (see discussion in
Section~3) tells us that
\be
\label{contour}
\oint_C P_{hh}(z)\ dz = 0\ .
\ee
Along the circle at infinity, with $z=Re^{i\theta}$, noting that
$P_{hh}\sim 1/(R e^{i\theta})$ we get a constant contribution:
\be
\oint_{C_\infty} P_{hh}(z)\ dz\simeq \int_0^{2\pi}
\frac{iRe^{i\theta}\ d\theta}{R e^{i\theta}} = 2i\pi\ .
\ee
The integral along the real axis is
\be
\label{axisint}
\oint_{C_{pole}} P_{hh}(z)\ dz + \int_{m_g^2}^\infty ds
[P_{hh}(s+i\epsilon)-
P_{hh}(s-i\epsilon)]\ ,
\ee
where $C_{pole}$ is an infinitesimal contour encircling clockwise the real
pole (at
$z=m_h^2$). The integral around this pole is evaluated using the theorem
of
residues and gives
\be
\oint_{C_{pole}} P_{hh}(z)\ dz = -2i\pi \left.\frac{1}{{\cal
D}'(s)}\right|_{s=m_h^2}= -2i\pi \frac{1}{K^2(m_h^2)}\ ,
\ee
so that one can also write
\be
\oint_{C_{pole}} P_{hh}(z)\ dz =
-2i\pi \int_0^{m_g^2}
\frac{1}{K^2(s)}\delta(s-m_h^2)\ ds\ =-2i\pi \int_0^{m_g^2} \rho_{hh}(s)\
ds\ .
\ee
For the second piece in (\ref{axisint}) we use (\ref{rhodef}) to write
$P_{hh}(s+i\epsilon)=-i\pi\rho_{hh}(s)$. Then, notice that for this
particular $P_{hh}(z)$ we also have (this is not always the case)
$P_{hh}(s-i\epsilon)=i\pi\rho_{hh}(s)$. Putting all pieces together,
(\ref{contour}) leads to
\be
\int_0^\infty \rho_{hh}(s)\ ds = 1\ .
\ee
This is the correct normalization of the spectral function for a stable
state.


\begin{thebibliography}{10}

%\cite{Patt:2006fw}
\bibitem{portal}
  B.~Patt and F.~Wilczek,
  %``Higgs-field portal into hidden sectors,''
  [hep-ph/0605188].
  %%CITATION = HEP-PH/0605188;%%
%\cite{Georgi:2007ek}
\bibitem{Georgi}
  H.~Georgi,
  %``Unparticle Physics,''
  Phys.\ Rev.\ Lett.\  {\bf 98} (2007) 221601
  [hep-ph/0703260]; 
%%CITATION = PRLTA,98,221601;%%
%\cite{Georgi:2007si}
%  H.~Georgi,
  %``Another Odd Thing About Unparticle Physics,''
  Phys.\ Lett.\  B {\bf 650} (2007) 275
  [hep-ph/0704.2457].
  %%CITATION = ARXIV:0704.2457;%%
% 
\bibitem{DEQ} 
%\bibitem{Delgado:2007dx}
  A.~Delgado, J.~R.~Espinosa and M.~Quir\'os,
  %``Unparticles-Higgs Interplay,''
  JHEP {\bf 0710} (2007) 094
  [hep-ph/0707.4309].
% 
%\cite{Delgado:2008rq}
\bibitem{DENQ}
  A.~Delgado, J.~R.~Espinosa, J.~M.~No and M.~Quir\'os,
  %``The Higgs as a Portal to Plasmon-like Unparticle Excitations,''
  [hep-ph/0802.2680].
  %%CITATION = ARXIV:0802.2680;%%
%
%\cite{Fox:2007sy}
\bibitem{shirman}
 P.~J.~Fox, A.~Rajaraman and Y.~Shirman,
  %``Bounds on Unparticles from the Higgs Sector,''
 Phys.\ Rev.\  D {\bf 76} (2007) 075004
  [hep-ph/0705.3092];
  %%CITATION = ARXIV:0705.3092;
 M.~Bander, J.~L.~Feng, A.~Rajaraman and Y.~Shirman,
  %``Unparticles: Scales and High Energy Probes,''
  Phys.\ Rev.\  D {\bf 76} (2007) 115002
  [hep-ph/0706.2677];
  %%CITATION = ARXIV:0706.2677;%%
%%CITATION = JHEPA,0710,094;
%\cite{Cacciapaglia:2007jq}
  G.~Cacciapaglia, G.~Marandella and J.~Terning,
  %``Colored Unparticles,''
  JHEP {\bf 0801} (2008) 070
  [hep-ph/0708.0005].
  %%CITATION = JHEPA,0801,070;%%
%
%\cite{Cacciapaglia:2008ns}
\bibitem{AdSCFTUn}
  G.~Cacciapaglia, G.~Marandella and J.~Terning,
  %``The AdS/CFT/Unparticle Correspondence,''
  [hep-ph/0804.0424].
  %%CITATION = ARXIV:0804.0424;%%
%
\bibitem{FA}
P.W.~Anderson, 
%``Localized Magnetic States in Metals,''
Phys. Rev. {\bf 124} (1961) 41;
U.~Fano,
%``Effects on Configuration Interaction on Intensities and Phase Shifts,''
Phys. Rev. {\bf 124} (1961) 1866;
G.D.~Mahan,
{\em Many-Particle Physics}, Plenum Press, New York, 1990.
%
%\cite{Giudice:2000av}
\bibitem{gravis}
  G.~F.~Giudice, R.~Rattazzi and J.~D.~Wells,
  %``Graviscalars from higher-dimensional metrics and curvature-Higgs mixing,''
  Nucl.\ Phys.\  B {\bf 595} (2001) 250
  [hep-ph/0002178].
  %%CITATION = NUPHA,B595,250;%%
%
\bibitem{fano}
%\cite{Fano:1992ru}
  U.~Fano,
  %``A Common Mechanism Of Collective Phenomena,''
  Rev.\ Mod.\ Phys.\  {\bf 64} (1992) 313.
  %%CITATION = RMPHA,64,313;%%
%
\bibitem{Cheung}
%\cite{Cheung:2007ue}
K.~Cheung, W.~Y.~Keung and T.~C.~Yuan,
%``Collider signals of unparticle physics,''
  Phys.\ Rev.\ Lett.\  {\bf 99} (2007) 051803
[hep-ph/0704.2588].
%%CITATION = ARXIV:0704.2588;%
%
%\cite{Stephanov:2007ry}
\bibitem{deco}
  M.~A.~Stephanov,
  %``Deconstruction of Unparticles,''
  Phys.\ Rev.\  D {\bf 76} (2007) 035008
  [hep-ph/0705.3049];
  %%CITATION = ARXIV:0705.3049;%%
%\cite{Krasnikov:2007fs}
  N.~V.~Krasnikov,
  %``Unparticle as a field with continuously distributed mass,''
Int.\ J.\ Mod.\ Phys.\  A {\bf 22} (2007) 5117
  [hep-ph/0707.1419].
  %%CITATION = ARXIV:0707.1419;%%
%
\bibitem{limited}
B.~Gaveau and L.S.~Schulman,
%``Limited Quantum Decay,''
J.\ Phys.\ A:\ Math.\ Gen.\ {\bf 28} (1995) 7359.
%
%\cite{Jahnke:2006nj}
\bibitem{nucl}
  L.~Jahnke and S.~Leupold,
  %``Complete relativistic description of the N*(1520),''
  Nucl.\ Phys.\  A {\bf 778} (2006) 53
  [nucl-th/0601072].
  %%CITATION = NUPHA,A778,53;%%
%
\end{thebibliography}
\end{document}